\documentclass[twocolumn,
aps,nofootinbib,showpacs,showkeys,preprint
tightenlines
] {revtex4-1}

\usepackage{epsf,epsfig,subfigure,graphicx,amsmath,amssymb}
\usepackage{color}
\usepackage{float}
\newcommand{\dis}[1]{\begin{equation}\begin{split}#1\end{split}\end{equation}}

\newcommand{\ovH}{\hskip 0.05cm \overline{\hskip -0.03cm H}}

\newcommand{\ie}{{\it i.e.}\ }
\newcommand{\etal}{{\it et al.}}

\newcommand{\one}{{\bf 1}}

\newcommand{\fif}{{\bf 15}}

\newcommand{\six}{{\bf 6}}
\newcommand{\sixb}{\overline{\bf 6}}

\newcommand{\two}{{\bf 2}}

\newcommand{\Z}{\mathbb{Z}}

\begin{document}


\title{\large\bf  $Z'$ from SU(6)$\times$SU(2)$_h$ GUT, $Wjj$ anomaly and Higgs boson mass bound}

\author{Jihn E. Kim$^{a,b}$\email{jekim@ctp.snu.ac.kr} and Seodong Shin$^a$
\email{sshin@phya.snu.ac.kr}
}
\affiliation{
$^a$FPRD and Department of Physics and Astronomy, Seoul National University, Seoul 151-747, Korea\\
$^b$GIST College, Gwangju Institute of Science and Technology,  Gwangju 500-712, Korea \\
 }
\begin{abstract}
A general electroweak scale $Z'$ is applied in a supersymmetric SU(6)$\times$SU(2)$_h$ grand unification model, to have a $\Z_6$ for the hexality.
We briefly show that there cannot exist any baryonic U(1)$'_B$ in any subgroup of E$_6$.
Any effect that requires sizable $Z'$ couplings to quarks like the reported Wjj anomaly of CDF, if observed, implies a substantial $Z'$ coupling to leptons or Higgs doublets. The kinetic mixing considered in a supersymmetric model from E$_6$ is restricted by the gauge coupling unification and neutrino mixing. The mass of $Z'$ is strongly constrained by the electroweak $\rho_0$ parameter. We conclude that $Z'$ mass much above 10 TeV is favored by considering the neutrino mixing and proton decay constraint in supersymmetric models. In this sense, the CDF $Wjj$ anomaly cannot be fitted to any electroweak model descending from E$_6$. Furthermore, if $Z'$ is found at several hundred GeV, any grand unification group embedded in E$_6$ such as SU(6)$\times$SU(2), SO(10),  SU(5)$\times$U(1), SU(5), SU(4)$\times$SU(4), and SU(3)$^3$, needs fine-tuned gauge couplings. We also discuss the U(1)$'$ effect on the tree level mass of the lightest MSSM Higgs boson.
\end{abstract}

\pacs{ 12.10.Dm, 12.60.Cn, 12.60.Jv, 11.25.Mj}

\keywords{ $Z'$ gauge boson, $Wjj$ anomaly, Higgs boson mass, SU(6)$\times$SU(2)$_h$ GUT}
\maketitle


In the standard model (SM), there is one electroweak scale neutral gauge boson $Z$ \cite{Kim81}. Discovery of any new neutral gauge boson hints the existence of a bigger gauge group beyond SU(3)$\times$SU(2)$\times$U(1). This possibility is widely discussed in view of the CDF reports such as $Wjj$ final states having a bump in dijet invariant mass near 145 GeV \cite{CDF11}\footnote{It must be also noticed that there is no such bump in the recent D0 result with 4.3 fb$^{-1}$ \cite{Abazov:2011af} and the LHC result at the 1 fb$^{-1}$ \cite{lhcwjj} integrated luminosity. With these results, the modeling of the SM background can be important as some of the papers in \cite{CDF11}.}. The simplest extension is just assuming a new U(1)$'$ beyond the SM gauge group, which does not fix the U(1)$'$ quantum numbers except by the constraints from the anomaly freedom  \cite{KimShin11}.

On the other hand, if the extension beyond the SM is achieved in (semi-)simple gauge groups, then the gauge quantum numbers are not arbitrary but fixed for given representations. The most analyzed grand unification (GUT) groups for U(1)$'$ are SO(10) and E$_6$ GUTs \cite{Langacker08}. However, there is the notorious gauge hierarchy problem in GUTs. The supersymmetry (SUSY) model was suggested to solve this problem.  In addition, the doublet/triplet splitting problem is the most serious issue with the SUSY GUTs.

The doublet/triplet splitting problem in GUTs is surfaced as the $\mu$-problem \cite{KimNilles84} in the minimal SUSY SM (MSSM). There are several ways to solve the $\mu$-problem in some extension of the MSSM. In particular, the string solutions seem to be interesting because they touch upon all other plausible phenomenological aspects of the MSSM from the ultraviolet completed theory \cite{JHKim07}. For instance, a SUSY electroweak group SU(3)$_W\times$U(1) is exceptionally useful for obtaining one pair of Higgs doublets $H_u$ and $H_d$ in the MSSM, naturally solving the $\mu$-problem \cite{KimGMSBst}.

In this paper, we present a U(1)$'$ model from a SUSY SU(6)$\times$SU(2)$_h$ GUT. For the U(1)$'$phenomenology from E$_6$, the discussion from its subgroup SU(6)$\times$SU(2)$_h$ is as good as E$_6$ since their ranks are the same. From the chain of GUTs, ${\rm SU(5)}\subset {\rm SO(10)} \subset {\rm E}_6$, the discussion of SO(10) is included in the discussion of E$_6$. Note also that the extra $Z'$ from SO(10) is in fact the $Z'$ of $B-L$ which is not the one we try to introduce for the recent $Wjj$ anomaly. Any U(1)$'$ generator can be written as a linear combination of six E$_6$ Cartan generators or of six SU(6)$\times$SU(2)$_h$ Cartan generators. If  the SU(6) is taken as a GUT, the electroweak part is SU(3)$_W\times$U(1) \cite{KimSU6}. Its SUSY extension was obtained from the F-theory compactification of string \cite{KSChoiKim11}. Note, however, that in our U(1)$'$ discussion, the SU(6) GUT is not a necessity except from the proton stability condition. The representation of SU(6)$\times$SU(2)$_h$ can be shown as matrix elements on the plane without any attachment of U(1) quantum numbers. This is a nice feature to glimpse the $Z'$ quantum numbers, just by looking at the representation on the plane. From these representations, we will notice that there are only two neutral SM singlets $N$ and $N'$ which are the heavy neutrinos needed for the seesaw mechanism.
The chief motivation for the SUSY GUT group containing SU(6) is from the proton hexality condition that forbids proton decay operators of dimension-4 and dimension-5 terms \cite{Dreiner06}. The $R$-parity forbids the dimension-4 operator from the superpotential $u^cd^cd^c$ but allows the dimension-5 proton decay operator from the superpotential $qqq\ell$. The SU(5) GUT does not forbid this dimension-5 proton decay operator, and its coefficient is required to be as small as $0.995 \times 10^{-8}$ considering the limit of the proton decay to $K^+ \bar{\nu}$ in \cite{PData10}. The hexality of Ref. \cite{Dreiner06} is the product of $R$-parity and triality which forms a $\Z_6$. The operator $qqql$ is allowed by the $R$-parity but is forbidden by the triality since four triality nonsinglet fields (with $q$'s having the same triality) are multiplied. The reason SU(5) cannot accommodate the hexality is that it does not have a discrete subgroup $\Z_6$. On the other hand, GUTs containing SU(6) can have the $\Z_6$ discrete subgroup since the center of SU(6) is $\Z_6$.

The SU(6)$\times$SU(2)$_h$ SUSY model allows the following representations, e.g. for the first family,

\dis{
\fif_L &\equiv (\fif,\one)=\left(\begin{array}{cccccc}
0&u^c& -u^c &u&d&D \\ -u^c&0&u^c & u&d&D \\
u^c&-u^c&0 &u&d&D \\ -u&-u&-u&0 & e^c&H_u^{+} \\
-d&-d&-d& -e^c&0 & H_u^{0} \\
-D&-D&-D&-H_u^{+}  & -H_u^{0} &0 \\
\end{array}\right),\\
\\
\sixb_{\two,1} &\equiv (\sixb,\two^\uparrow)=\left(\begin{array}{c}
  {d^c}\\ { d^c}\\ { d^c} \\  -\nu_{e}  \\ {e} \\  { N}
\end{array}\right),\
\sixb_{\two,2} \equiv (\sixb,\two^\downarrow)=\left(\begin{array}{c}
  {D^c}\\  {D^c}\\  {D^c}\\   -H_d^0   \\  H_d^- \\  N'
\end{array}\right).\label{eq:SUsixGUT}
}
Note that we have not included any E$_6$ singlets\footnote{According to the recent LHC result with the 1 fb$^{-1}$ integrated luminosity, there are new strict lower bounds on the masses of colored exotics by the study of dijet resonances. The bounds are 2.91 TeV for the excited quarks, 3.21 TeV for axigluons, and 1.91 TeV for color octet scalars \cite{lhcexotic}.}.
The representations $\sixb_{\two,1}$ and $\sixb_{\two,2}$ form a doublet pair of the horizontal group SU(2)$_h$. Without the loss of generality, we choose $\sixb_{\two,1}$ as matter and $\sixb_{\two,2}$ as Higgs sextets.
We need three families and at least a vector-like pair $n_{(\sixb,\two)}$  and  $n_{(\six,\two)}$, which is responsible for the breaking SU(6) down to SU(5). Therefore, to allow for three chiral families and SU(6)$\to$SU(5) breaking, we assume $n_{(\sixb,\two)}=4$  and  $n_{(\six,\two)}=1$ \cite{KSChoiKim11}. By the vacuum expectation values (VEVs) of $\sixb_\two$ and $\six_\two$, SU(6)$\times $SU(2)$_h$ is broken down to SU(5)$\times $U(1)$'$.

By the VEV of the adjoint representation {\bf 35} of SU(6) or by the hyper-flux in F-theory \cite{Beasley09}, our interest is focused on a rank 6 group SU(3)$_c\times$SU(3)$_W\times$U(1)$\times$SU(2)$_h$.\footnote{Without confusion, we can use the GUT representations to simplify the notation at the scale where the broken group is effective.}
To break SU(3)$_W\times$U(1)$\times$SU(2)$_h$ down to the one including the rank 4 SM group SU(3)$_c\times$SU(2)$_W\times$U(1)$_Y$, we assign a GUT scale VEVs to $(\sixb,\two)$ and $(\six,\two)$, which reduce just one rank, and the low energy gauge group is SU(3)$_c\times$SU(2)$_W\times$U(1)$_Y\times$U(1)$'$. The tensor form of the representation $(\sixb,\two)$ is $\sixb^{\,\alpha}_i$ where $\alpha=1,2,\cdots,6$ and  $i=1,2$. The fields of Eq. (\ref{eq:SUsixGUT}) couple as
\dis{
f\,\fif_{\alpha\beta}\sixb^{\,\alpha}_i  \ovH^{\,\beta}_j\epsilon^{ij},\label{eq:Higgs62}
}
where we suppressed the family indices and used $\ovH$ as another $(\sixb,\two)$.
The group SU(2)$_W\times$U(1)$_Y\times$U(1)$'$ contains three diagonal generators, $Q_{\rm em}, Q_{Z}$, and $Y'$. The SM $Q_{\rm em}$ and $Q_{Z}$ are included in the GUT SU(6), representing the linear combinations of two SU(6) generators only in the vertical directions of  $(\six,\two)$:
$T_3=\rm diag.\,(0\,0\,0\,\frac{1}{2}\,\frac{-1}{2}\,0)$ and $Y=\rm diag.\,(\frac{-1}{3}\,\frac{-1}{3}\,\frac{-1}{3}\,\frac{1}{2}\,\frac{1}{2}\, 0)$.
The U(1)$'$ generator is a linear combination of two SU(6)$\times$SU(2)$_h$ diagonal generators in the vertical and horizontal directions of  $(\six,\two)$: $Y_{\rm SU(6)}$ and $X_3$.

Let the gauge bosons corresponding to $T_3, Y$, and $Y'$ be $A_\mu^3, B_\mu$ and $C_\mu$(with coupling $g^{\prime\prime}$), respectively . Below, we will present the form of $Y'$. The mass eigenstates are defined as the photon $A_\mu$, $Z_\mu$-boson and $Z'_\mu$-boson. In this extended weak interaction model,\footnote{Even if only $Z'$ survives down to the electroweak scale, we call it a new weak interaction model.} we define a new weak mixing angle $\sin^2\varphi = g^{\prime\prime\,2}/(g^2+g^{\prime\, 2}+g^{\prime\prime\, 2})$ in addition to $\sin^2\theta_W=g^{\prime\,2}/(g^2+g^{\prime\, 2})\simeq 0.23$ of the SM. The diagonal gauge bosons of SU(2)$_W\times$U(1)$_Y\times$U(1)$'$ ($A^3_\mu, B_\mu, C_\mu$) are related to the mass eigenstate gauge bosons($A_\mu, Z_\mu, Z'_\mu$) by an orthogonal matrix,
\dis{
\left(\begin{array}{c}
A^3_\mu \\ B_\mu\\ C_\mu \end{array}\right)
=\left(\begin{array}{ccc}
s_\theta &-c_\theta c_\varphi & c_\theta s_\varphi\\ c_\theta & s_\theta c_\varphi&  -s_\theta s_\varphi \\
0&s_\varphi& c_\varphi
\end{array}\right)\left(\begin{array}{c}
A_\mu \\ Z_\mu\\ Z'_\mu \end{array}\right).\label{eq:weakphi}
}
where $s_\theta=\sin\theta_W$ and $s_\varphi=\sin\varphi$, and similarly for the cosines. The gauge boson masses depend on the $Y'$ quantum numbers of Higgs fields, which will be discussed below.

Below, we prove the no-go theorem for a gauged U(1)$'_B$ from E$_6$ and its consequence on the $Z$ boson and the lightest CP-even Higgs boson masses.
Finally, we comment on the possibility of obtaining  SU(6)$\times$SU(2)$_h$ from the ultraviolet completed superstring.

\vskip 0.2cm
 {\it On gauged U(1)$'_B$ and leptophobic U(1)$'$ from E$_6$}:
The chiral representation {\bf 27} of E$_6$ is split into $(\fif,\one)$ and $(\sixb,\two)$ of Eq. (\ref{eq:SUsixGUT}). Rank 6 E$_6$ has six diagonal generators: $F_3$ and $F_8$ of SU(3)$_c$, $T_3$ of SU(2)$_W$, $Y$ of U(1)$_Y$, $Y_{\rm SU(6)}\equiv Y_6$, and $X_3$. In any subgroup of E$_6$, the diagonal generators are linear combinations of these . Therefore, without loss of generality, we consider the baryon number as a linear combination of $Y, Y_6$ and $X_3$. To have an $R$-parity, we include a global U(1)$_R$ symmetry and consider the following U(1)$'_B$
\dis{
B=aY+b Y_6+cX_3+d R
}
where $Y_6\equiv Y_{\rm SU(6)}= \rm diag.\,(\frac{-1}{6}\,\frac{-1}{6}\,\frac{-1}{6}\,\frac{-1}{6}\,\frac{-1}{6}\, \frac{5}{6} )$ is for the representation $\six$,  $X_3\equiv \rm diag. (\frac12\,-\frac12)$ is the third generator of SU(2)$_h$, and $R$ is the U(1)$_R$ charge. The $R$-symmetry is broken at the high-energy scale, we set $d=0$ and the resulting $B$ would be a gauge group generator.

If $B$ is a good symmetry, leptons and Higgs fields should carry vanishing $B$. In addition, $u^c$ and $d^c$ must carry the same $B$, which is opposite to that of the quark doublet $(u,d)_L$.
The required conditions of leptons and Higgs fields are
\dis{
 e^c&:~\quad \quad a-\frac13 b\quad \quad =0\\
 (\nu,e)&:~-\frac12 a+\frac16 b+\frac12 c =0\\
  H_d &:~-\frac12 a+\frac16 b-\frac12 c =0\\
  H_u &:~+\frac12 a+\frac23 b\quad\quad =0
\label{eq:condition}
}
which cannot be satisfied unless $a=b=c=d=0$.
Therefore, it is not possible to have a gauged U(1)$'_B$ as a subgroup of E$_6$.
\footnote{The U(1)$'_B$ model such as \cite{u1b} is not originated from E$_6$.}

But, not requiring a strict baryon number, it is possible to consider a useful nonbaryonic U(1)$'$ from E$_6$. It is the so-called  ``leptophobic" that leptons do not have the U(1)$'$ interaction, \ie  no  U(1)$'$ charges for $ (\nu,e)$ and $e^c$.
Since $H_d$ has the same charge as $ (\nu,e)$ in the MSSM, $H_d$ should not carry U(1)$'$ charge. Therefore, by adopting the first three conditions only in Eq. (\ref{eq:condition}), a solution $a=\frac13 b$ and $c=0$ is obtained.
Making $H_u$ not the complex-conjugate of $H_d$, \ie going beyond the SM, we note that this can be realized in the so-called two Higgs doublet model with $H_u$ carrying a nonvanishing U(1)$'$ charge.
In this case, the diagonal entries of the leptophobic U(1)$'$ charge is
\dis{
Y'_{\rm lp-phob}=\frac{5}{6} \Big(\frac{-1}{3},\frac{-1}{3},\frac{-1}{3},0,0,1 \Big).\label{eq:Yleptopho}
}
There are two ways to realize this $Y'_{\rm lp-phob}$. One is to introduce a VEV of the adjoint {\bf 78} of E$_6$, and the other is by considering the kinetic mixing between $B_{\mu}$ and $C_{\mu}$ in our model. Needing the adjoint representation {\bf 78} is very much involved in the orbifold construction \cite{Kakushadze96}, but is easily achievable in an F-theory construction.

Even without the VEV of {\bf 78}, the kinetic mixing between $B_{\mu}$ and $C_{\mu}$ has been considered with the branching E$_6 \to$ SO(10)$ \times$ U(1)$_{\psi} \to$ SU(5) $\times$ U(1)$_{\chi} \times$ U(1)$_{\psi}$, including the running of the gauge couplings \cite{kinetic}. However, this case needs an extreme fine-tuning between masses of the split multiplet members
to obtain such a large mixing. In addition, the leptophobia obtained by the kinetic mixing with coefficient 1/3 is not achieved if one requires an anomaly free model where the gauge couplings of the SM and U(1)$'$ are perturbative and unify at the GUT scale \cite{noexact}.

On the other hand, to give singlet neutrino masses, $N$ of Eq. (\ref{eq:SUsixGUT}) should develop a VEV, implying that $Y'_{\rm lp-phob}$ is broken at the heavy neutrino mass scale. Therefore, the exact leptophobic $Z'$ from E$_6$ should be very heavy to induce the neutrino mixing, which cannot explain the recent $Wjj$ anomaly of CDF. In this sense, our U(1)$'$ is introduced not to be leptophobic by assigning no charge to $N$ so that the $Z'$ charge $Y' = X_3 + \frac35 Y_6$.

Therefore, let us consider $Z'$ from E$_6$, coupling to baryons, couples to leptons as well.

\vskip 0.2cm
{\it $Z'$ and Higgs boson masses}:
If only the third family members have VEVs, without loss of generality, we can choose $\langle N\rangle=V_{\rm heavy}$ and $\langle N'\rangle=0$.  We also introduce at least one vector-like representations $(\six,\two)$ and $(\sixb,\two)$ as in Ref. \cite{KSChoiKim11}. Since there is no parameter space where the leptonic U(1)$'$ currents are negligible, the high precision NC experiments and the LEP II data for nonvanishing $Z'-$lepton coupling stringently restrict the $Z'$ mass.

The neutral fields carrying nonvanishing U(1)$'$ charges are $H^0_{u,d}, \nu, $ and $N'$. For $Z'$ to survive down to the electroweak scale, $N'$ should not develop a superheavy VEV above the electroweak scale. However, for the neutrino oscillation, we also need them to be heavy \cite{KSChoiKim11} with mass larger than $10^{10}$ GeV. The $N'$ Majorana mass can be generated by $\frac{1}{M_P}(\six,2)(\six,2)(\sixb,2)(\sixb,2)$ as in Ref. \cite{KSChoiKim11} by the VEV $\langle (\six,\two)\rangle\to \langle \ovH^{\,6}_1\rangle= \langle N\rangle$. Using Eq. (\ref{eq:Higgs62}), the Dirac mass between $\nu$ and $N'$ is generated by $\langle \ovH_u\rangle$. These lead to the seesaw mechanism for neutrino masses. The $H^0_{u,d}$ and $\nu_e$ fields at the electroweak scale carry nonvanishing U(1)$'$ charges. For the $R$-parity conservation, $\nu_e$ is not required to break U(1)$'$. In addition, three $N'$ fields survive down to the electroweak scale because $N'$ and $\overline{N}^{\,\prime}$ fields in four $\sixb$'s and one $\six$ remove only one heavy Dirac neutrino field, viz. by $\frac{1}{M_P}(\six,\two) (\six,\two)(\sixb,\two)(\sixb,\two)$,
\dis{
\begin{array}{ccccccc}
& N'_1~ & N'_2~& N'_3~ & N'_4~& \overline{N}\,'&
\end{array}&\\
\begin{array}{c}
N'_1 \\ N'_2\\ N'_3 \\ N'_4\\ \overline{N}\,' \end{array}
\left(\begin{array}{ccccc}
0 &0 & 0 & 0 & M_1\\
0 &0 & 0 & 0 & M_2\\
0 &0 & 0 & 0 & M_3\\
0 &0 & 0 & 0 & M_4\\
M^*_1 & M^*_2 & M^*_3 & M^*_4 &  M
\end{array}\right)&.\label{eq:NbarNmass}
}
where the masses $M$ and $M_{1-4}$ are at the intermediate scale. The electroweak singlet neutrinos are called $N_i^{\,\prime}\,(i=1,2,3)$ again. The VEVs of $H_u, H_d$, and $N_i^{\,\prime}\,(i=1,2,3)$ break the U(1)$'$ symmetry at the electroweak scale. Since we look for the parameter space, where $g^{\prime\prime}$ is smaller than the SU(2)$_W$ coupling $g$, the contribution of the $H_u$ and $H_d$ VEVs to $M_{Z'}$ is smaller than their contribution to the $W$ boson mass. Therefore, there must be a TeV scale VEV(s) of $N_i^{\,\prime}\,(i=1,2,3)$ to make $Z'$ as heavy as 150 GeV, if the CDF $Wjj$ rate is attributed to $Z'$. This additional VEVs are free parameters to tune the $Z'$ mass. Since $Y_6=\frac25$ and $-\frac25$ for $H_u$ and $H_d$, respectively, the $Z-Z'$ mass matrix becomes as following. Here $s_\varphi,\, c_\varphi$ are defined in Eq. (\ref{eq:weakphi}), $t_\varphi=s_\varphi/c_\varphi,\, \tan^2\gamma\equiv X^2/V^2,\, G\equiv\sqrt{g^2+g^{\prime\,2}},\, V^2\equiv v_u^2+v_d^2$ and $X^2$ is the contribution from the VEVs of $N'$ fields.

\begin{figure}
\includegraphics[width=6.cm]{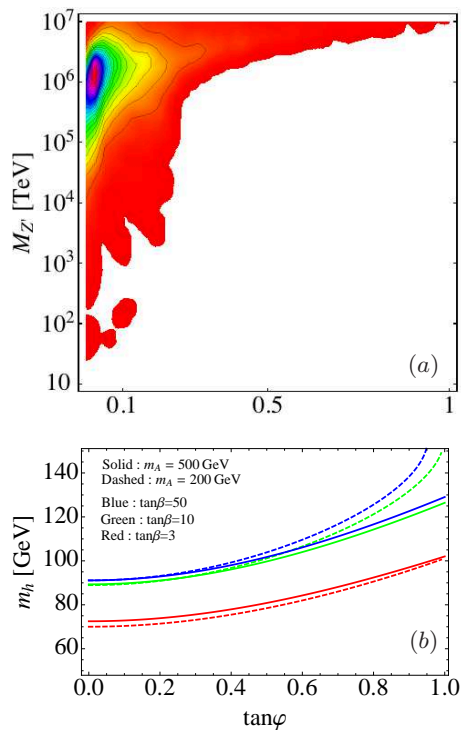}
\caption{Masses of ($a$) $Z'$ and ($b$) the lightest CP-even Higgs $m_h$  as functions of $\tan\varphi$.
The total number of data points is $14,235$, and the region with no data points is white. From the red color (grey color of the outmost contour), the colors are separated by the density of points, increment of ten for each step.
} \label{fig:mh}
\end{figure}

\begin{widetext}
\dis{
M&^2_{ZZ'}=\frac{G^2V^2}{4}\left( \begin{array}{cc} c^2_\varphi+\frac85 s^2_\varphi+\left(\frac{16}{25}+4\tan^2\gamma \right)t^2_\varphi s^2_\varphi\,,  & -\frac15 c_\varphi s_\varphi+4t_\varphi  s^2_{\varphi}\tan^2\gamma -\frac{4}{25}t_\varphi  s^2_{\varphi}\\[0.5em]
 -\frac15 c_\varphi s_\varphi+4t_\varphi  s^2_{\varphi}\tan^2\gamma -\frac{4}{25}t_\varphi  s^2_{\varphi}\,,
& \left(\frac{16}{25}+4\tan^2\gamma \right)s^2_\varphi-\frac35 s^2_\varphi
\end{array}\right) \label{eq:Massgauge}
}
\end{widetext}
From Eq. (\ref{eq:Higgs62}), we note that the $N'$ VEVs break the $R$-parity.
If any four fields of $N'$ in Eq. (\ref{eq:NbarNmass}) does not develop a VEV, we can consider the limit $X^2\ll v_d^2\ll v_u^2$, \ie $\tan\gamma\simeq 0$. Generally, this case leads to $M_{Z'}$ smaller than $M_Z$, and
we are left with a large  $\tan\gamma$ case. Not to be conflicted with the $R$-parity problem, a  large  $\tan\gamma$  must be provided by the heavy pair of $N'$ and $\overline{N}^{\,\prime}$. The VEV of the 4$^{\rm th}~N'$ combines the lepton doublets with the superheavy $H_u$. This case is not ruled out obviously. Even for this large  $\tan\gamma$, the $\rho$ parameter constrains the allowed mass of $Z'$.  For this study, we satisfy the electroweak neutral current (NC) parameter $\rho_0=1.0004^{+0.0029}_{-0.0011}$ with the $2\sigma$ limit, which has no meaningful bound on the Higgs mass \cite{PData10}.
We show the allowed $\tan\varphi$ and $M_Z^{\prime}$ in the region $g^{\prime\prime\,2}<g^{\prime\,2}$ in Fig. \ref{fig:mh}($a$), from which we note that the heavy $Z'$ much above 10 TeV is favored in the region $M_{Z'}>M_Z$.
Adding to this, there are more constraints such as $Z$ boson decay width, $e^- e^+ \to W^- W^+$, etc. However, as seen in Fig. \ref{fig:mh}($a$), the constraint on $\rho_0$ parameter provides strong enough conclusion to constrain the viable $Z'$ parameters. Therefore, we leave the analysis by considering other experimental constraints to our next work.

The VEVs of $H_u$ and $H_d$ are related to the Higgs boson masses and can raise the upper bound of the lightest Higgs boson mass of the MSSM,  even before including the radiative corrections \cite{Haber91}. If $Z'$ is present and the Higgs doublets, $H_u$ and $H_d$, carry nonzero U(1)$'$ charges, the lightest CP even Higgs boson mass bound is changed. Our interest based on the $\rho$ parameter constraint is Case $H2$ of Eq. (20) of Ref. \cite{Civetic97}.  In the limit $\tan\beta\equiv v_u/v_d \to\infty$, it can be shown succinctly. In the MSSM, we have $v_u^2=8\mu_u^2/G^2$ and hence $m_h^2\simeq M_Z^2$ for $v_d\ll v_u$. With the $Y_6$ contribution in the $D$-term potential, $-\mu_u^2 H_u^\dagger H_u+\frac14G^2 (1 +\frac{16}{25}\tan^2\varphi)( H_u^\dagger H_u)^2 +(H_d~{\rm terms}),$ we obtain $m_H^2=2\mu_u^2+\cdots$and $M_Z^2=\frac14 G^2V^2$. Then, the $Z$ boson mass has the same expression as in the MSSM, but the relation between the VEV $v_u^2$ and $\mu_u^2$ is changed to  $v_u^2=8\mu_u^2/G^2(1 +\frac{16}{25} \tan^2\varphi)$ if we neglect the Higgsino mixing $\mu$ term and obtain the tree level upper bound on the Higgs boson mass as, $ {m^2_H} ={M_Z^2}(1+\frac{16}{25}\tan^2\varphi)$.

However, in this limit, the bound represents the heavier CP even Higgs since the real part of $H_d^0$ is massless, if we neglected its mass parameter $\mu_d^2$. If $Z'$ is decoupled at high energy for $X\gg V$, still the MSSM Higgs boson masses are strongly affected. The reason is that the lightest MSSM Higgs boson mass encodes the quartic couplings or the gauge symmetry (whether it is broken or not).   The lightest Higgs mass bound is not protected by the decoupling theorem since the dimensionless quartic couplings are renormalized only logarithmically.

In the limit $\tan\beta \equiv v_u/v_d\to\infty$, the pseudoscalar mass $m_A$ goes to zero as commented above. So, we must consider a finite  $\tan\beta$ case, \ie for nonzero $v_d$ and also for the Higgsino mixing term $\mu$ \cite{KimNilles84} to make the pseudoscalar heavy. In this case, we consider the following $2\times 2$ CP even Higgs mass matrix
\begin{widetext}
\dis{
M^2_{\rm CP~even}&=\left( \begin{array}{cc}
m_A^2c_\beta^2+M_Z^2s_\beta^2+\frac{8}{25}M_Z^2 t^2_\varphi(4s_\beta^2-1),& -(m_A^2+M_Z^2[1+\frac{16}{25}t^2_\varphi])c_\beta s_\beta \\ [0.5em]
-(m_A^2+M_Z^2[1+\frac{16}{25} t^2_\varphi])c_\beta s_\beta ,&  m_A^2s_\beta^2+M_Z^2c_\beta^2+\frac{8}{25}M_Z^2 t^2_\varphi(4c_\beta^2-1) \end{array}\right) \label{eq:CPevenMass}
}
where we parameterized the $B_\mu$ term as the pseudoscalar mass $m^2_A$, $t^2_\varphi$ is defined in Eq. (\ref{eq:Massgauge}), and $s_\beta(c_\beta)$ is $\sin\beta(\cos\beta)$.

Equation (\ref{eq:CPevenMass}) leads to the following eigenvalues for the lighter and the heavier CP even Higgs fields, for $m_A>M_Z$
\dis{
m_{h,H}^2=&\frac12 \left(m_A^2+M_Z^2 +\frac{16}{25}t^2_\varphi\right) \mp\frac12
\left(\left[m_A^2+M_Z^2 +\frac{16}{25} t^2_\varphi\right]^2\right.\\
 &\quad \left.-4\cos^2 2\beta \left[m_A^2M_Z^2-\frac{8}{25}M_Z^2 t^2_\varphi(-3m_A^2+M_Z^2+ \frac{24}{25}M_Z^2 t^2_\varphi) \right]\right)^{1/2}.\label{eq:hmass}
}
\end{widetext}

The condition to obtain positive $m_h^2$ is subtle because the term in the second line of Eq. (\ref{eq:hmass}) cannot be too large. It severely depends on the CP-odd Higgs mass $m_A$. The dependence of $m_h$ on $\tan\varphi$ is depicted in Fig. \ref{fig:mh}($b$) for a few values of $m_A$ and $\tan\beta$. The $\tan\beta$ dependence converges in the large $\tan\beta$ region.

\vskip 0.3cm
{\it Comment related to F-theory}:
The above SU(6)$\times$SU(2)$_h$ model can be obtained from the F-theory construction \cite{Vafa96,ChoiKS10}. In this construction, we first obtain a visible six dimensional (6D) GUT group, which is then broken to the four dimensional (4D) SM group by fluxes. To obtain the 6D GUT group, one should consider the holonomy groups, the continuous and the discrete ones. The SU(3)$_\perp$ holonomy alone does not specify the 6D group completely. The additional information on the discrete holonomy is needed to choose one of the rank 6 groups amomg E$_6$, SU(6)$\times$SU(2) and SU(3)$^3$. To guarantee the proton longevity, forbidding the dimension-5 operators, the hexality has been proposed.  For a natural hexality, a visible sector should have $\Z_6$ \cite{Dreiner06,Forste10,KSChoiKim11}. The centers of SU($N$) and ${\rm E}_6$ are $\Z_N$ and $\Z_3$, respectively. So, we rule out 6D E$_6$ but require 6D SU(6) part. Therefore, to introduce $\Z_6$ holonomy of SU(6)$\times$SU(2), we look for $\Z_6$ holonomy of SU(3)$_\perp$. Since the center of SU(3)$_\perp$ is $\Z_3$, we need additional $\Z_2$, which is possible if the holonomy of the instanton is from the Belavin \etal~ type instanton \cite{Belavin75}. Then, it is possible to have a 6D GUT group SU(6)$\times$SU(2).\\

\vskip 0.2cm
{\it Conclusions}:

We analyzed the electroweak scale $Z'$ in the context of a supersymmetric U(6)$\times$SU(2)$_h$ grand unification model, which provides a $\Z_6$ for the hexality to make it safe from the dangerous proton decay. Motivated from the recent CDF result on the Wjj excess around 150 GeV, we analyzed the possibility of constructing a leptophobic $Z'$ from our model. However, we briefly showed that there cannot exist any baryonic U(1)$'_B$ in any subgroup of E$_6$. Aside from U(1)$'_B$, the leptophobic $Z'$ model from a supersymmetric E$_6$ is usually constructed through the kinetic mixing historically. Such mixing demands a large mixing coefficient 1/3 which can arise from fine-tuned relations between the masses of the split multiplet members. It is also not achieved if one requires an anomaly free model where the gauge couplings of the SM and U(1)$'$ are perturbative and unify at the GUT scale according to other research. Such a scenario is also constrained by the neutrino mixing.

Analyzing the electroweak $\rho_0$ parameter, the mass of our $Z'$ is favored to be above 10 TeV by considering the neutrino mixing and proton decay constraint in supersymmetric models. In this sense, the CDF $Wjj$ anomaly cannot be fitted to any electroweak model descending from E$_6$. We also discussed the U(1)$'$ effect on the tree level mass of the lightest MSSM Higgs boson and the F-theory construction to obtain our supersymmetric  SU(6)$\times$SU(2)$_h$ model.


\vskip 0.2cm
\noindent {\bf Acknowledgments}: {We thank K.-S. Choi for helpful discussions. This work is supported in part by the National Research Foundation  (NRF) Grant No. 2005-0093841 funded by the Korean Government (MEST).}




\begin{thebibliography}{99}

\def\prp#1#2#3{Phys.\ Rep.\ {\bf #1} (#3) #2}
\def\rmp#1#2#3{Rev. Mod. Phys.\ {\bf #1} (#3) #2}
\def\anrnp#1#2#3{Annu. Rev. Nucl.
Part. Sci.\ {\bf #1} (#3) #2}
\def\npb#1#2#3{Nucl.\ Phys.\ {\bf B\,#1} (#3) #2}
\def\plb#1#2#3{Phys.\ Lett.\ {\bf B\,#1} (#3) #2}
\def\prd#1#2#3{Phys.\ Rev.\ {\bf D\,#1}, #2 (#3)}
\def\prl#1#2#3{Phys.\ Rev.\ Lett.\ {\bf #1} (#3) #2}
\def\jhep#1#2#3{J. High Energy Phys.\ {\bf #1} (#3) #2}
\def\jcap#1#2#3{J. Cos. Astropart. Phys.\ {\bf #1} (#3) #2}
\def\zp#1#2#3{Z.\ Phys.\ {\bf #1} (#3) #2}
\def\epjc#1#2#3{Euro. Phys. J.\ {\bf #1} (#3) #2}
\def\ijmp#1#2#3{Int.\ J.\ Mod.\ Phys.\ {\bf #1} (#3) #2}
\def\mpl#1#2#3{Mod.\ Phys.\ Lett.\ {\bf #1} (#3) #2}
\def\apj#1#2#3{Astrophys.\ J.\ {\bf #1} (#3) #2}
\def\nat#1#2#3{Nature\ {\bf #1} (#3) #2}
\def\sjnp#1#2#3{Sov.\ J.\ Nucl.\ Phys.\ {\bf #1} (#3) #2}
\def\apj#1#2#3{Astrophys.\ J.\ {\bf #1} (#3) #2}
\def\ijmp#1#2#3{Int.\ J.\ Mod.\ Phys.\ {\bf #1} (#3) #2}
\def\apph#1#2#3{Astropart.\ Phys.\ {\bf B\,#1}, #2 (#3)}

\bibitem{Kim81} S.~L.~Glashow, Nucl.~Phys.~{22}{579}{1961}; J. E. Kim, P. Langacker, M. Levine, and H. H. Williams,
\rmp{53}{211}{1981}.

\bibitem{CDF11} T. Aaltonen \etal(CDF Collaboration), arXiv:1104.0699. For the SM explanation, see, Z. Sullivan and A. Menon, arXiv:1104.3790 [hep-ph];  T. Plehn and M. Takeuchi,
  arXiv:1104.4087 [hep-ph].

\bibitem{Abazov:2011af}
  V.~M.~Abazov {\it et al.}  [D0 Collaboration],
  Phys.\ Rev.\ Lett.\  {\bf 107}, 011804 (2011)
  [arXiv:1106.1921 [hep-ex]].

\bibitem{lhcwjj} ATLAS Collaboration, ATLAS-CONF-2011-097.

\bibitem{KimShin11} J.~Erler, \npb{586}{73}{2000} [hep-ph/0006051]. For a recent discussion with more references, see, J. E. Kim, M.-S. Seo and S. Shin, \prd{83}{036003}{2011} [arXiv:1010.5123 [hep-ph]].

\bibitem{Langacker08} For a review, see, P. Langacker, \rmp{81}{1199}{2009} [ arXiv:0801.1345 [hep-ph]].

\bibitem{KimNilles84} J. E. Kim and H. P. Nilles, \plb{138}{150}{1984}.

\bibitem{JHKim07} J. E. Kim, J.-H. Kim and B. Kyae,   \jhep{0706}{034}{2007} [hep-ph/0702278]; O. Lebedev, H. P. Nilles, S. Raby, S. Ramos-Sanchez, M. Ratz, and P. K.S. Vaudrevange, and A. Wingerter, \prd{77}{046013}{2008} [arXiv:0708.2691 [hep-th]]. For GUTs, see, J. E. Kim and B. Kyae, \npb{770}{47}{2007} [hep-th/0608086]; K. J. Bae, J. H. Huh, J. E. Kim, B. Kyae, and R. D. Viollier, \npb{817}{58}{2009} [arXiv:0812.3511]; J.-H. Huh, J. E. Kim, and B. Kyae, \prd{80}{115012}{2009} [arXiv: 0904.1108 [hep-ph]].

\bibitem{KimGMSBst} J. E. Kim, \plb{656}{207}{2007} [arXiv: 0707.3292].

\bibitem{KimSU6} J. E. Kim, \plb{107}{69}{1982}. See, also, M. Abud, F. Buccella, H. Ruegg, and C. A. Savoy \plb{67}{313}{1977}; P. Langacker, G. Segr\`e and A. Weldon, \plb{73}{87}{1978}; H. Georgi and A. Pais, \prd{19}{2746}{1979}.

\bibitem{KSChoiKim11} K.-S. Choi and J. E. Kim, \prd{83}{065016}{2011} [arXiv: 1012.0847[hep-ph]].

\bibitem{Dreiner06} H. K. Dreiner, C. Luhn and M. Thormeier, \prd{73}{075007}{2006} [arXiv:hep-ph/0512163].

\bibitem{PData10} K. Nakamura \etal (Particle Data Group), J. Phys. {\bf G\,37}, 075021 (2010).

\bibitem{lhcexotic} ATLAS collaboration, ATLAS-CONF-2011-095.

\bibitem{Beasley09} C. Beasley, J. J. Heckman and C. Vafa,  \jhep{01}{058}{2009} [arXiv:0802.3391 [hep-th]] and  \jhep{01}{059}{2009} [arXiv:0806.0102 [hep-th]]; J. Marsano, N. Saulina, and S. Schafer-Nameki, \jhep{08}{046}{2009} [arXiv:0906.4672 [hep-th]].

\bibitem{u1b}  C.~D.~Carone and H.~Murayama, \prl{74}{3122}{1995} [arXiv:hep-ph/9411256];
  C.~D.~Carone and H.~Murayama, \prd{52}{484}{1995}  [arXiv:hep-ph/9501220].

\bibitem{Kakushadze96} For an SO(10) adjoint in the fermionic construction, see, Z. Kakushadze and S. H.-H. Tye, \prl{77}{2612}{1996} [hep-th/9605221].

\bibitem{kinetic}  K.~S.~Babu, C.~F.~Kolda and J.~March-Russell, \prd{54}{4635}{1996}  [arXiv:hep-ph/9603212];
  K.~R.~Dienes, C.~F.~Kolda and J.~March-Russell, \npb{492}{104}{1997}  [arXiv:hep-ph/9610479];
  K.~S.~Babu, C.~F.~Kolda and J.~March-Russell, \prd{57}{6788}{1998}   [arXiv:hep-ph/9710441];
  J.~Erler, P.~Langacker, S.~Munir and E.~R.~Pena, \jhep{0908}{017}{2009}  [arXiv:0906.2435 [hep-ph]].
  For left-right symmetry,
  F.~del Aguila, G.~A.~Blair, M.~Daniel and G.~G.~Ross, \npb{283}{50}{1987}.

\bibitem{noexact}
  T.~G.~Rizzo, \prd{59}{015020}{1999}   [arXiv:hep-ph/9806397];
  K.~Leroux and D.~London, \plb{526}{97}{2002}  [arXiv:hep-ph/0111246].

\bibitem{Haber91} Y. Okada, M. Yamaguchi and T. Yanagida, Prog. Theor. Phys. {\bf 85}, 1 (1991); \plb{262}{54}{1991}; J. Ellis, G. Ridolfi, and F. Zwirner, \plb{257}{83}{1991}; H. E. Haber and R. Hempfling, \prl{66}{1815}{1991}; R. Barbieri and M. Frigeni, \plb{258}{395}{1991}.

\bibitem{Civetic97} M. Cvetic, D. A. Demir, J. R. Espinosa, L. Everett, and P. Langacker, \prd{56}{2861}{1997} [arXiv:hep-ph/9703317].

\bibitem{Vafa96} C. Vafa, \npb{469}{403}{1996} [arXiv:hep-th/9602022].

\bibitem{ChoiKS10} The construction method is summarized in, K.-S. Choi, \npb{842}{1}{2011} [arXiv:1007.3843 [hep-th]], and references therein.


\bibitem{Forste10} S. F\"orste, H. P. Nilles, S. Ramos-Sanchez, and P. K. S. Vaudrevange, \plb{693}{386}{2010} [arXiv:1007.3915 [hep-ph]].

\bibitem{Belavin75}  A. A. Belavin, A. Polyakov, A. Schwartz, and Y. Tyupkin, \plb{59}{85}{1975}.


\end{thebibliography}
\end{document}